\documentclass[sigconf]{acmart} 
\AtBeginDocument{%
  }

\copyrightyear{2026}
\acmYear{2026}
\setcopyright{cc}
\setcctype{by}
\acmConference[Submitted to ICMI '26]{Proceedings of the 28th International Conference on Multimodal Interaction}{October 5--9, 2026}{Napoli, Italy}
\acmBooktitle{Proceedings of the 28th International Conference on Multimodal Interaction (ICMI '26), October 5--9, 2026, Napoli, Italy}

\usepackage{graphicx}
\usepackage{hyperref}
\usepackage{cleveref}
\usepackage{booktabs}
\usepackage{multirow}
\usepackage{subcaption}

\usepackage{fontawesome5} 

\setcopyright{none}
\begin{document}

\title{IntentVLM: Open-Vocabulary Intention Recognition through Forward–Inverse Modeling with Video-Language Models}

\author{Hamed Rahimi, Clémence Grislain, Adrien Jacquet Crétides, Olivier Sigaud, Mohamed Chetouani\\
Institute of Intelligent Systems and Robotics (ISIR),  Sorbonne University\\ Paris, France\\
  \small\texttt{firstname.lastname}@isir.upmc.fr}

\renewcommand{\shortauthors}{Rahimi et al.}

\begin{abstract}

Improving the effectiveness of human–robot interaction requires social robots to accurately infer human goals through robust intention understanding. This challenge is particularly critical in multimodal settings, where agents must integrate heterogeneous signals including text, visual cues to form a coherent interpretation of user intent. This paper presents \textit{IntentVLM}, a novel two-stage video-language framework designed for open-vocabulary human intention recognition. The approach is inspired by forward-inverse modeling in cognitive science by decomposing intention understanding into goal candidate generation followed by structured inference through selection, effectively reducing hallucinations in latent reasoning. Evaluated on the IntentQA and Inst-IT Bench datasets, IntentVLM achieves state-of-the-art results with up to 80\% accuracy, notably surpassing the baseline performance by 30\% and matches human performance. Our findings demonstrate that this structured reasoning approach enhances open-vocabulary intention understanding without catastrophic forgetting, offering a robust foundation for human-centered robotics.
\end{abstract}



\begin{teaserfigure}
\centering
  \includegraphics[width=0.8\linewidth]{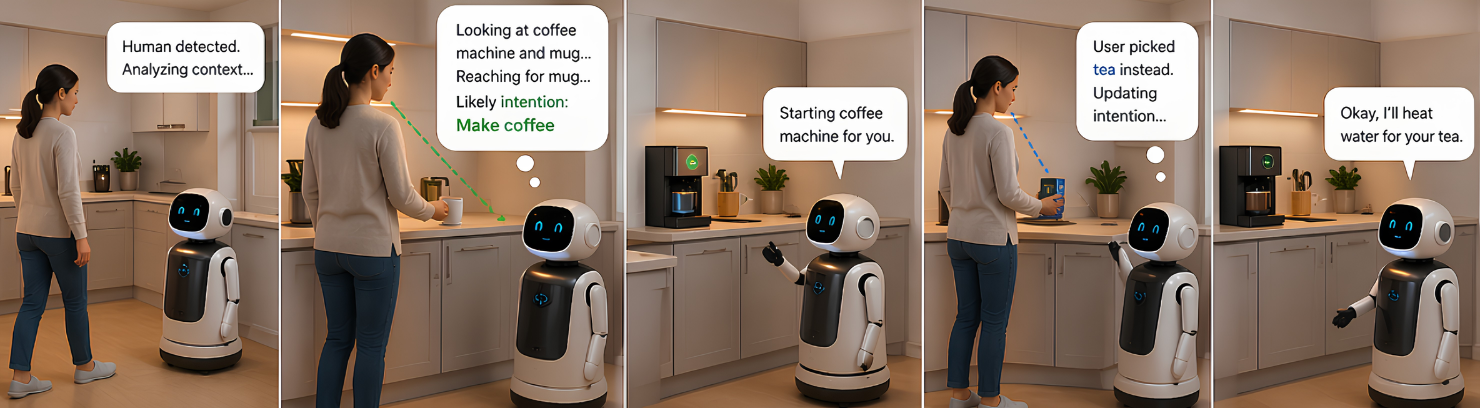}
  \caption{IntentVLM for open-vocabulary human intention recognition from video streams. Given a continuous video stream of a human in a kitchen environment, IntentVLM integrates multimodal visual cues — scene context, object interactions, and body movements — to infer free-form user intentions without relying on a predefined action vocabulary. The model reasons over temporal video evidence to produce natural language intention descriptions ("Make coffee", "Heat water for tea"), demonstrating the capacity of video-language understanding to ground open-vocabulary human goals in real-world robotic settings.}
  \label{fig:teaser}
\end{teaserfigure}

\received{April 2026}

\maketitle

\faGithub  ~~ \href{https://github.com/hamedR96/IntentVLM}{Code and Data} 

\section{Introduction}
As robots increasingly transition from controlled industrial environments into everyday \textit{proactive} robot human-centered spaces~\cite{van2024proactive}, their ability to understand and respond appropriately to human behavior becomes critically important\cite{malecot2026harmoni}. In such settings, effective interaction requires more than simply perceiving observable actions, meaning robots must also reason about the underlying human intentions that drive those actions. Human intention refers to a mental state representing a goal-directed commitment to performing a particular action~\cite{brune2007mental, Chetouani2026}. Within the framework of mental state attribution, intentions function as internal states that guide and motivate behavior, enabling an observer to interpret why an agent behaves in a particular way. For example, as shown in \Cref{fig:teaser}, when a person enters a reception area and begins scanning the room (e.g. looking at people, objects, or walls) their behavior cannot be misinterpreted as casual curiosity, but in reality, this pattern more commonly signals an underlying intention to seek guidance or assistance \cite{ziemke2020understanding}. Recognizing such subtle behavioral cues is essential to accurately infer user intent and respond appropriately.

\begin{figure*}
    \centering
  \includegraphics[width=0.8\linewidth]{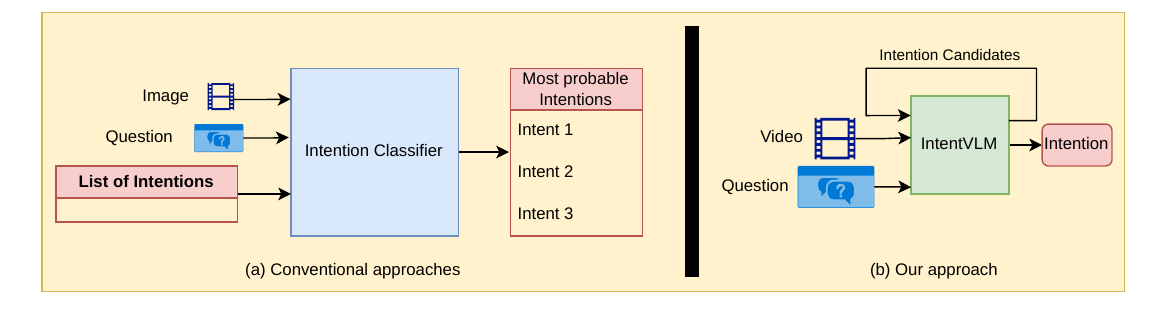}
  \caption{Closed-vocabulary solutions vs our approach. Conventional approaches formulate intention recognition as closed-set classification over a predefined label space given visual inputs. Inspired by forward–inverse modeling in cognitive science, our method removes this requirement by decomposing the task into (i) goal candidate generation and (ii) structured selection via a video–language model. This two-stage formulation reduces hallucinations in latent reasoning and supports open-vocabulary intention prediction.}
  \label{fig:arch}
\end{figure*}

While intention is often discussed in the context of human cognition, it is not exclusive to humans. Animals exhibit goal-directed behavior, and autonomous robotic systems act according to internal models and task-oriented objectives \cite{thellman2022mental}. In this broader view, an agent’s behavior can be understood through the interplay of beliefs (probabilistic representations of the world), goals (desired states to be achieved), intentions (commitments to particular courses of action), and policies (action-selection mechanisms) \cite{Chetouani2026}. Consequently, intention recognition can be defined as the process of inferring an agent’s latent intentions (i.e., its committed plans or policies) by reasoning over observable actions in relation to underlying beliefs, goals, and environmental constraints. Rather than focusing solely on what an agent does, intention recognition seeks to explain why those actions are performed and to predict future behavior accordingly \cite{hoffman2024inferring}.

Despite its importance, intention recognition remains a challenging problem \cite{jain2019probabilistic}. The primary difficulty arises from the fact that intentions are inherently unobservable. Unlike physical actions, which can be directly perceived through sensory input, intentions exist only as latent mental states that must be inferred indirectly \cite{hoffman2024inferring}. Observers therefore infer an agent’s intention, which is its underlying goal-directed commitment, by integrating behavioral patterns, contextual information, prior knowledge, and communicative cues~\cite{cretides2026encoding}. Furthermore, intention expression is inherently multimodal \cite{trick2019multimodal}. Humans convey intent through interactions with the environment with a combination of visual and linguistic signals. Another challenge stems from the imperfect relationship between actions and intentions \cite{nguyen2011capir}. Not every intention culminates in an observable action, nor does every action faithfully reveal the agent’s underlying intention. A goal denotes the desired outcome an individual seeks to achieve, whereas intention captures the specific commitment or plan to act toward that outcome. In practice, these constructs may diverge: an agent may intend to perform a beneficial action yet fail due to external constraints, or may execute an apparently helpful behavior while pursuing a self-serving objective~\cite{grislain2025failsense}. Such misalignment between goals, intentions, and observable actions introduces substantial ambiguity in behavior interpretation, thereby making robust intention inference a central challenge for robotic systems.

A large body of research has addressed intention recognition from multiple perspectives \cite{hoffman2024inferring}, including Bayesian inverse planning \cite{buyukgoz2021proactive}, probabilistic reasoning \cite{VanHorenbeke2021Activity}, plan recognition \cite{Kautz1986plan}, and learning-based approaches \cite{sukthankar2014pair}. However, as shown in \Cref{fig:arch} many existing methods simplify the problem by restricting possible intentions to a predefined set, often represented by a limited vocabulary of intention labels~\cite{zhao2025deep}. While such formulations simplify learning and evaluation, they significantly restrict the expressiveness of intention modeling. Human intentions frequently involve nuanced psychological states, social goals, or compound motivations that cannot be easily represented by a single discrete label \cite{buyukgoz2021proactive}. Additionally, a large portion of prior work relies primarily on static image-based perception, which limits the system's ability to capture temporal dynamics and contextual evolution present in real-world scenes. As a result, integrating intention recognition into complex, temporally evolving visual scenes remains an open challenge.

\begin{figure}
    \centering
    \includegraphics[width=1\linewidth]{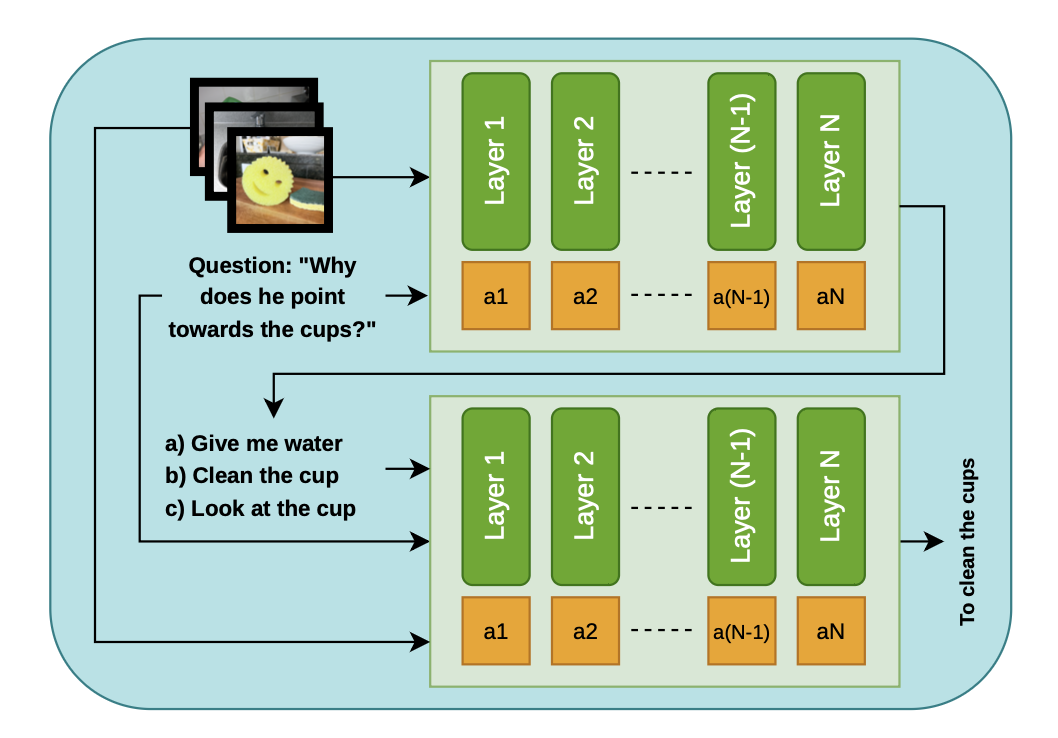}
    \caption{Architecture of IntentVLM. We cast the model as a cognitive framework based on forward–inverse reasoning, implemented with two frozen video–language modules augmented by trained LoRA adapters. The first module generates a diverse set of candidate intentions for each video–query pair, while the second acts as an expert that evaluates and selects the most likely intention. The model is trained on egocentric video data, where a robot seeks to infer human intentions from its embodied viewpoint. The LoRA adapters are optimized using a cross-entropy loss computed between the ground-truth labels (intention candidates for the first module and the true human intention for the second module) and the tokens generated by the video-language model.}
    \label{fig:det_arch}
\end{figure}

To address these limitations, as shown in \Cref{fig:det_arch}, we introduce \textit{IntentVLM}, a video foundation model \cite{madan2024foundation} designed for open-ended human intention understanding. We formulate intention recognition as a cognitive modeling problem grounded in forward–inverse reasoning \cite{ho2022cognitive}, and instantiate it as a video–language reasoning task using two vision–language models (VLMs) \cite{zhang2024vision}. Given a video of a scene and a textual query about the intention of an observed agent, the system outputs a textual description of the inferred intention. Our framework comprises two complementary modules. The first generates a diverse set of candidate intentions conditioned on the video context, while the second evaluates and ranks these candidates via multimodal reasoning to select the most plausible one. This two-stage design enables broader exploration of the intention space while reducing hallucinations and improving reliability.

Extensive experiments demonstrate that the proposed approach provides more accurate and contextually grounded intention predictions compared to baseline video foundation models. Moreover, the modular design enables improved computational efficiency and reduced model complexity while maintaining strong reasoning capability. These results establish structured intention recognition as a key component for enabling video-language foundation models to support more socially aware and cognitively capable robotic systems.
\section{Related Work}
%


Intention recognition, closely related to goal recognition \cite{lesh1995gr} and plan recognition \cite{Kautz1986plan}, refers to the process of inferring an agent’s goals and plans from observed behavior. Intention, goal, and plan are closely related notions, which helps explain why similar methodological frameworks are often employed to infer them. However, this similarity is not only conceptual but also practical: all three tasks rely on interpreting observable actions to reconstruct latent structures, often under uncertainty and partial observability, leading to shared modeling assumptions and inference techniques. Despite these commonalities, intention recognition can be distinguished as a broader process, as it aims to infer an actor’s underlying mental states and high-level objectives as they are manifested through actions \cite{bratman1989}. In this sense, it subsumes aspects of both goal and plan recognition while extending beyond them.

In the literature, developed intention recognition systems are broadly categorized into probabilistic and logic-based approaches~\cite{VanHorenbeke2021Activity}. A foundational framework for formalizing this problem is Bayesian Inverse Planning, which frames intention recognition as the inversion of an agent’s decision-making process~\cite{charniak1991pr}. By formulating the task as a Markov Decision Process, probabilistic methods~\cite{Baker2009Action,Ramrez2009PlanRA, ramrez2010pr} define intent as a latent variable that maximizes the likelihood of observed actions. This approach is particularly effective at managing real-world uncertainty and the inherent "non-rationality" of human behavior. However, these models often face scalability challenges due to state-space explosion. To mitigate this, Schrempf et al.~\cite{Schrempf2007Tractable} utilize expert knowledge to derive reduced state spaces for hidden variables within Bayesian networks.

In contrast, logic-based methods rely on plan representation languages and domain-independent rules to deduce intent through formal reasoning~\cite{sukthankar2014pair, masters2019}. While these approaches offer high expressivity and transparency, they often struggle to model the stochastic nature of human actions or sub-optimal decision-making \citep{dreyfus2007}. To bridge this gap, hybrid frameworks have been developed to combine the deductive strength of logic with the robustness of Bayesian inference~\cite{buyukgoz2021proactive}, aiming to provide systems that are both interpretable and capable of handling environmental noise. Prior work in cognitive science has framed intention recognition as an instance of inverse modeling, tightly coupled with forward predictive processes. Early accounts emphasized that predictive (forward) models of others’ behavior rely on prior inference over communicative or action intentions, rather than purely stimulus-driven prediction \citep{deRuiter2013}. More recent computational approaches formalize this relationship using Bayesian inverse planning, where observers infer latent goals by assuming approximately rational action, and combine this with forward simulation to predict future behavior \citep{Baker2009,Qian2021}. This forward–inverse loop has been further developed in frameworks such as inverse reinforcement learning, which infer underlying reward structures or shifting intentions from observed trajectories \citep{Ng2000,HadfieldMenell2016}. Integrative perspectives \citep{HoGriffiths2022} unify these approaches under a broader view of Theory of Mind as inverse inference over decision-making processes, supported by forward generative models for prediction~\cite{abrini2025proceedings}.

The emergence of Foundation Models has shifted the paradigm of embodied AI from narrow task-specific solvers to general-purpose reasoning agents \cite{huang2022icml, huang2022inner, khan2025fmreview}. Unlike classical models that require hard-coded rules, Large Language Models (LLMs) possess commonsense priors developed through massive-scale pre-training \cite{brown2020fewshotlearners}. This allows for "zero-shot" inference of high-level objectives, where a model can hypothesize an agent's intent by grounding observed actions in its broad world knowledge \cite{wei2022llm}.
Vision-Language Models (VLMs) extend this reasoning capability to the physical world by aligning visual perceptions with linguistic semantics \cite{li2025vlmreview,rahimi2025user,rahimi2025user2}. Early alignment efforts, such as contrastive frameworks like CLIP \cite{clip} introduced open-vocabulary recognition, enabling systems to identify novel objects and actions without predefined categories \cite{gao2024vlm}. This is critical for intention recognition in unconstrained environments, where the set of possible human goals is theoretically infinite. Modern modular architectures, such as Qwen3-VL \cite{bai2025qwen3}, further refine this by using lightweight adapters to bridge frozen visual encoders with LLM reasoning cores~\cite{rahimi2025demographic, rahimi2025reasoning}. This evolution ensures that intention recognition is no longer confined to predefined state spaces but can instead leverage the world knowledge of foundation models to interpret human behavior in unconstrained settings. Prior work studied LLMs for intent understanding in social interactions \cite{sap2022tom} and robotic QA \cite{chiu2021}, as well as RL-based goal inference in dynamic and collaborative settings \cite{nageris2024draco,lin2025reinforcement}. More recent approaches use LLMs to infer goals from observed actions or dialogue \cite{ying2024goma,zhang2025combo,wan2025fiser}. While early multimodal methods exist \cite{li2021visualnlp}, VLMs now enable grounding language into actions and tracking long-term intent; e.g., LIT \cite{huang2024lit} models behavior from video, and OV-MER \cite{Lian2024OVMER} adds open-vocabulary emotion recognition.

Building on these developments, our work addresses a key limitation in existing approaches: the lack of structured reasoning mechanisms for dynamic multimodal, open-vocabulary intention understanding. While recent LLM- and VLM-based methods demonstrate strong zero-shot capabilities, they often rely on implicit, single-pass inference, which can lead to hallucinations and inconsistent goal predictions in complex, real-world scenarios. 
Our design enables more robust integration of heterogeneous signals such as video and language, while preserving the flexibility of foundation models in open-world settings. By combining the expressivity of VLMs with an explicit reasoning structure, our approach bridges the gap between classical probabilistic inference and modern foundation model capabilities, yielding improved accuracy and human-level performance on challenging benchmarks.

\section{Method}

\subsection{Problem Definition}
\label{sec:preliminaries}

Intentions are latent components of humans' cognitive state that govern goal-directed behavior. Although intentions are not directly observable, they can be inferred indirectly through the analysis of observed actions and multimodal behavioral cues. We therefore formulate \textit{intention recognition} as the problem of inferring human’s underlying goal and intention conditioned on partial observations of their behavior.

Consider an observer monitoring a human interacting with an environment over time. The observer perceives a sequence of state-action pairs forming an observation trajectory:
\begin{equation}
O_{0:T} = \langle S_0, A_0,~ S_1, A_1,~ \dots,~ S_T, A_T \rangle,
\end{equation}
where $S_t \in \mathcal{S}$ denotes the observed environmental state at time $t$, $A_t \in \mathcal{A}$ represents the low-level action performed by the agent. 
The sequence captures the observable evolution of the interaction between humans and their environment.

However, in realistic scenarios the observer does not have direct access to these structured variables. Instead, the observer receives multimodal sensory inputs such as visual observations and linguistic queries. These observations provide only partial information about the underlying state, action, and goal structure governing the agent’s behavior.


We model the latent \textit{intention} guiding the behavior of humans as a hidden variable $I_t \in \mathcal{I}$ representing the internal objective of the agent at time $t$. While goal and intention are conceptually distinct in the cognitive literature, in this work we assume that the observed agent is neither deceptive nor pursuing unattainable goals. Under these constraints, the agent's observable behavior reliably reflects both their goal and their intention, allowing us to treat the two concepts as interchangeable for the purposes of this work. In our setting, the observing robot interacts with two primary modalities: an egocentric visual stream obtained from onboard cameras and a textual query that may originate from a human user or from the robot’s internal reasoning module. Let $V_O$ denote the observed video sequence and $Q_O$ denote the textual query describing the intention-related question. These modalities offer uncertain evidence of the human’s behavior and surrounding environment, yet neither the agent’s goal $G_t$ nor latent intention $I_t$ is directly observable. The robot must therefore reason under partial observability.

We hypothesize that explicitly modeling an intermediate goal inference step improves intention understanding, drawing inspiration from forward-inverse modeling frameworks in cognitive science, where the brain is thought to first simulate plausible future states before reasoning backward to infer the underlying intent of an action~\cite{HoGriffiths2022}. Rather than predicting intention directly from multimodal observations, we first infer a set of plausible goal candidates and subsequently reason over them to infer the most likely intention. This hierarchical formulation encourages the model to interpret observations in terms of goal-oriented behavior before making a higher-level cognitive inference about intention — mirroring the forward model's role in anticipating outcomes and the inverse model's role in attributing intent. Furthermore, grounding this process in a a video-language model naturally extends the framework to an open-vocabulary setting, enabling the system to express and reason over intentions beyond a fixed label space, capturing the context-dependent diversity of human goals in unconstrained environments. Such a decomposition allows the system to leverage richer contextual cues and reduces ambiguity in the final prediction.

\subsection{IntentVLM}

To investigate this hypothesis, we propose \textit{IntentVLM}, a video-language model for multimodal intention reasoning. The framework comprises two sequential components—a \textit{goal candidate generator} and an \textit{intention inference module}—that operate in two stages:

\paragraph{Stage 1: Proposal}
The first module takes as input the observed video $V_O$ and the textual query $Q_O$, and produces a set of $K$ plausible goal candidates, that explain the observed behavior, in form of multiple choices. Intuitively, these goals correspond to intermediate hypotheses about the human's underlying intent in the scene.

We model this component as a conditional generative model built upon a video-language model parameterized by $\phi$. Formally, the model generates a set of candidate goals conditioned on the observed modalities:
\begin{equation}
 \mathcal{G}_t  = F_{\phi}(V_O, Q_O),
\end{equation}
where $ \mathcal{G}_t =\{G_1, \dots, G_K\}$ denotes the set of generated candidate goals, which may belong to either a fixed vocabulary or an open vocabulary. Here, $F_{\phi}$ represents the parameterized video-language model that produces candidate goals based on the visual observations and the textual query.

\paragraph{Stage 2: Selection}
The second module performs intention inference by evaluating candidate goals and selecting the one that best explains the observed behavior. As mentioned before, we treat this selected goal as the inferred human intention, leveraging the close relationship between goals and intentions in goal-directed behavior: intentions reflect a commitment to achieve a desired outcome, while goals specify that outcome. Under this view, identifying the goal that most plausibly accounts for the observed actions provides a practical proxy for intention inference. Concretely, the model ranks candidate goals based on their consistency with the visual evidence and contextual information, and selects the one that maximizes explanatory alignment with the agent’s behavior.

Formally, let $\mathcal{G}_t = \{G_t^1, G_t^2, \dots, G_t^i\}$ denote the candidate goal set generated by the first module. In practice, we retain only the top-5 candidates from the first module, ensuring that the second-stage scoring remains tractable. The second module evaluates each candidate conditioned on the observed modalities and assigns them a likelihood score. We parameterize this scoring function using a model with parameters $\theta$:

\begin{equation}
P_\theta(G_t^k \mid V_O, Q_O, \mathcal{G}_t),
\end{equation}

where $G_t^k$ represents a candidate goal from the generated set. The inferred intention is obtained by selecting the goal with the highest posterior probability:

\begin{equation}
I_t = \arg\max_{G_t^k \in \mathcal{G}_t} P_\theta(G_t^k \mid V_O, Q_O, \mathcal{G}_t).
\end{equation}

The selected goal $G_t^{*}$ is therefore treated as the estimated intention $I_t$ of the agent. This formulation allows the model to explicitly reason over multiple plausible interpretations before committing to the most consistent explanation of the observed behavior.


\begin{figure}
    \centering
    \includegraphics[width=1\linewidth]{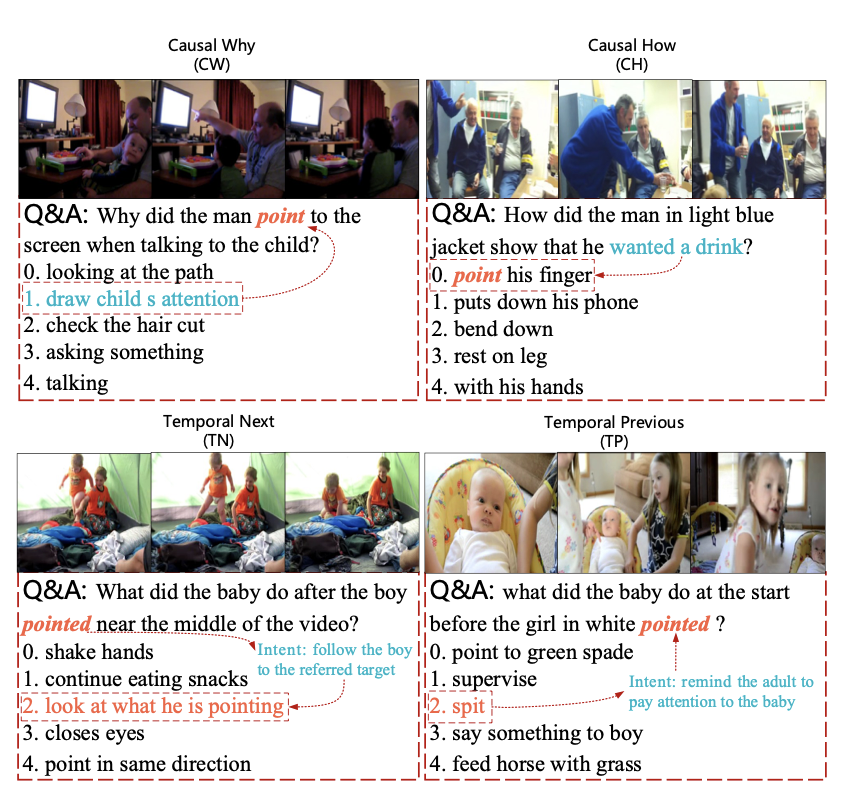}
    \caption{IntentQA Benchmark \cite{li2023intentqa}. Illustrative examples of the four QA types in our training dataset. CW: a man points at a screen to guide a child’s attention. CH: a pointing gesture expresses the intent to drink. TN: a boy’s pointing causes the baby to look in that direction. TP: a girl’s pointing is triggered by the baby’s action. The red box indicates the correct answer.}
    \label{fig:intqa}
\end{figure}

\begin{figure}
    \centering
    \includegraphics[width=1\linewidth]{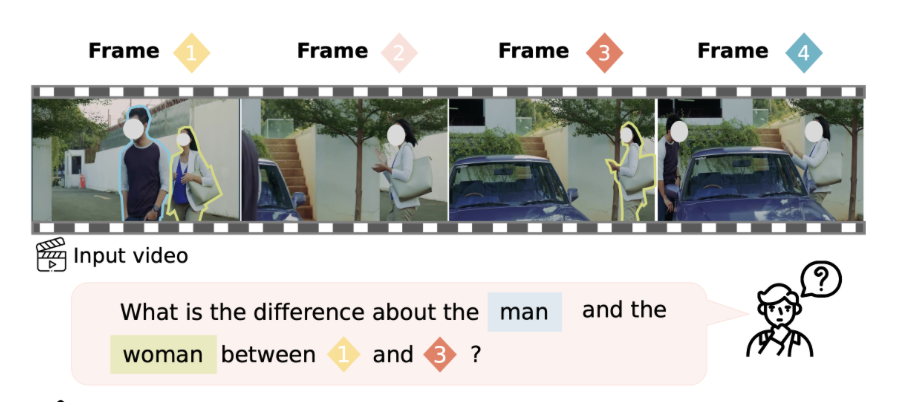}
    \caption{Inst-IT Benchmark \cite{peng2024inst}. Each sample consists of temporally grounded frame-level annotations with instance IDs, a coherent video-level description, and instance-focused QA pairs. The annotations capture fine-grained attributes, interactions, and temporal changes of instances across frames, enabling explicit instance-level reasoning.}
    \label{fig:insbench}
\end{figure}

\section{Experiments}

To evaluate the effectiveness of the proposed framework, we train two LoRA adapters corresponding to the two stages of IntentVLM: (i) goal candidate generation and (ii) intention inference via goal selection. Our experimental study is designed to answer the following key questions: (1) How do video-language models perform on intention recognition compared to existing baseline methods? (2) Does an open-vocabulary formulation combined with a two-stage (proposal and selection) framework improve performance? (3) Does training for intention reasoning improve generalization, or does it negatively affect visual understanding capabilities? (4) How does the scale of the underlying video-language model impact performance?

To systematically investigate these questions, we design an experimental framework comprising two benchmarks. The first benchmark addresses the initial two research questions, focusing on (i) plain intention recognition and (ii) open-vocabulary comparison. The second benchmark evaluates the model’s susceptibility to catastrophic forgetting, as well as its scalability, particularly in terms of the performance-to-size trade-off.

\begin{table*}[t]
\centering

\begin{tabular}{lc|cc|cc|cc|cc}
\toprule
 \multirow{2}{*}{Model} & \multirow{2}{*}{Text Rep.}
& \multicolumn{2}{c|}{CW}
& \multicolumn{2}{c|}{CH}
& \multicolumn{2}{c|}{TP\&TN}
& \multicolumn{2}{c}{Total} \\
\cmidrule(lr){3-4} \cmidrule(lr){5-6} \cmidrule(lr){7-8} \cmidrule(lr){9-10}
 & & Val. & Test & Val. & Test & Val. & Test & Val. & Test \\
\midrule
 EVQA$^*$     & GloVe & 25.99 & 25.92 & 37.43 & 34.54 & 28.00 & 25.52 & 28.38 & 27.27 \\
 CoMem$^*$   & GloVe & 31.56 & 30.00 & 35.63 & 28.69 & 28.57 & 28.95 & 31.46 & 29.52 \\
 HGA$^*$    & GloVe & 29.45 & 32.00 & 35.03 & 30.64 & 29.71 & 31.05 & 30.43 & 31.54 \\
 HME$^*$      & GloVe & 30.97 & 34.40 & 35.33 & 34.26 & 34.29 & 29.14 & 32.53 & 33.08 \\
 HQGA$^*$     & GloVe & 32.49 & 33.20 & 38.32 & 34.26 & 34.48 & 36.57 & 33.95 & 34.21 \\
\midrule
CoMem$^*$   & BERT  & 46.75 & 47.68 & 57.49 & 54.87 & 41.71 & 39.05 & 47.21 & 46.77 \\
HGA$^*$  & BERT  & 43.54 & 44.88 & 56.89 & 50.97 & 42.48 & 39.62 & 45.45 & 44.61 \\
 HME$^*$    & BERT  & 46.50 & 46.08 & 51.20 & 54.32 & 44.76 & 40.76 & 46.82 & 46.16 \\
 HQGA$^*$     & BERT  & 45.91 & 48.24 & 57.78 & 54.32 & 44.76 & 41.71 & 47.55 & 47.66 \\
 VGT$^*$      & BERT  & 50.46 & 51.44 & 55.99 & 55.99 & 48.19 & 47.62 & 50.78 & 51.27 \\
 Blind GPT$^*$ & -- & -- & 52.16 & -- & 61.28 & -- & 43.43 & -- & 51.55 \\
\midrule
 VideoQA$^*$ w/o GPT & BERT & 52.57 & 55.28 & 57.47 & 61.56 & 46.10 & 47.81 & 51.71 & 54.50 \\
 VideoQA$^*$         & BERT & -- & 58.40 & -- & 65.46 & -- & 50.48 & -- & 57.64 \\
\midrule
 \textbf{IntentVLM (ours)}  & -- 
& \textbf{85.30} & \textbf{84.10} 
& \textbf{83.80} & \textbf{88.60 }
& \textbf{75.30} & \textbf{83.95 }
& \textbf{79.93} & \textbf{85.15} \\
\midrule
Human$^*$ & -- & -- & 77.76 & -- & 80.22 & -- & 79.05 & -- & 78.49 \\
\bottomrule
\end{tabular}
\caption{Comparison of different models and Total accuracy (\%). $^*$ Values are reported by \cite{li2023intentqa}.}
\label{tab:main_results}
\end{table*}

\subsection{Datasets}

We evaluate our model on two multimodal benchmarks. The first focuses on intention recognition and understanding. The second assesses whether the model avoids catastrophic forgetting while preserving global scene understanding in complex tasks. In particular, we examine its performance on instance-level understanding and reasoning. Instance detection identifies what objects are present and where, while intention detection infers the underlying goal or purpose behind an observation.

\paragraph{IntentQA.}
The first dataset, \textit{IntentQA} \cite{li2023intentqa}, is designed specifically for intention recognition in video contexts. Each sample in the dataset consists of a video, an intention-related question, a set of candidate answers, and a ground-truth label. Formally, each data point can be represented as:
\begin{equation}
x = (V, Q, \mathcal{A}, a^*),
\end{equation}
where $V$ denotes the input video, $Q$ is the intention-related query, $\mathcal{A} = \{a_1, a_2, \dots, a_5\}$ represents a set of five candidate answers, and $a^* \in \mathcal{A}$ is the correct answer. The dataset is divided into four categories based on the type of reasoning  required: \textit{Causal What (CW):} understanding what intention explains an observed action, \textit{Causal How (CH):} reasoning about how an intention leads to a specific outcome, \textit{Temporal Next (TN):} predicting future actions based on inferred intention, \textit{Temporal Past (TP):} inferring past intentions from observed outcomes. An example of this dataset is provided in \Cref{fig:intqa}.

\begin{figure*}
    \centering
    \begin{subfigure}{0.6\linewidth}
        \centering
        \includegraphics[width=\linewidth]{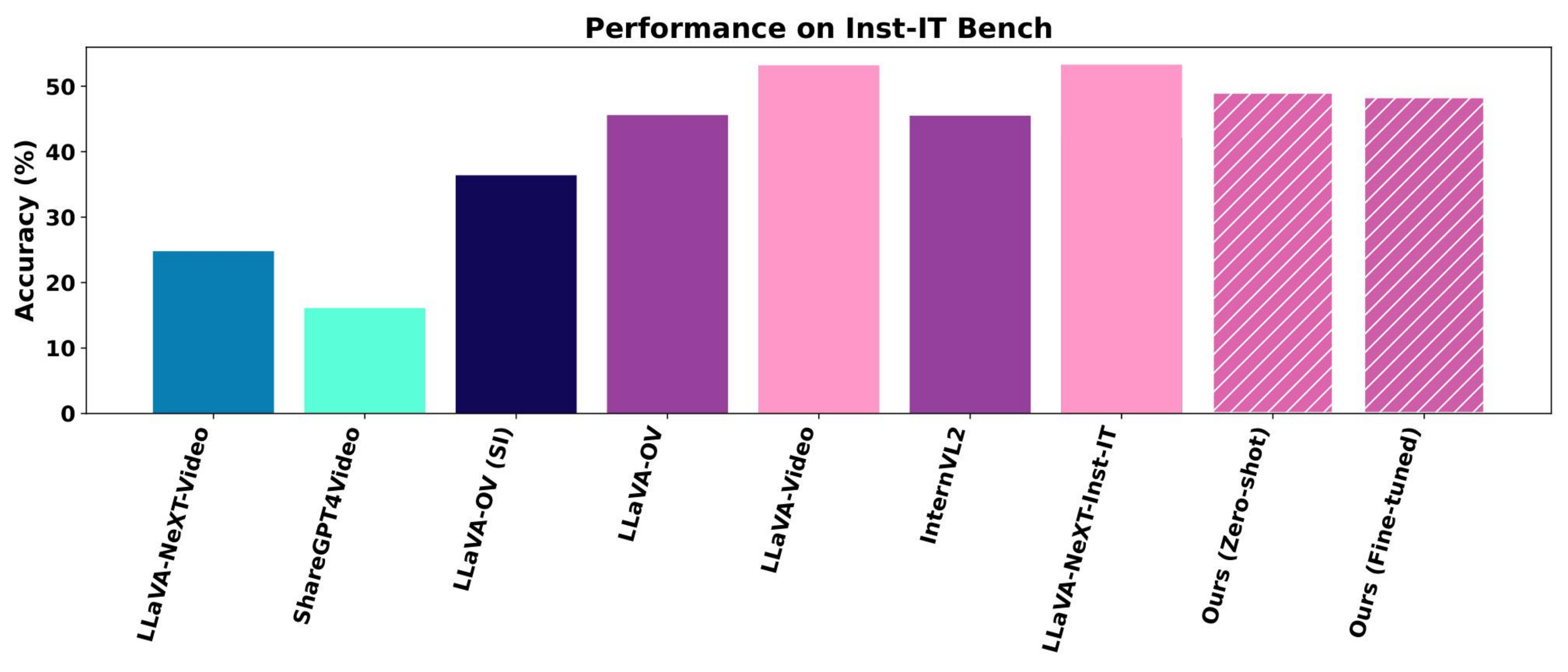}
        \caption{}
    \end{subfigure}
    \hfill
    \begin{subfigure}{0.37\linewidth}
        \centering
        \includegraphics[width=\linewidth]{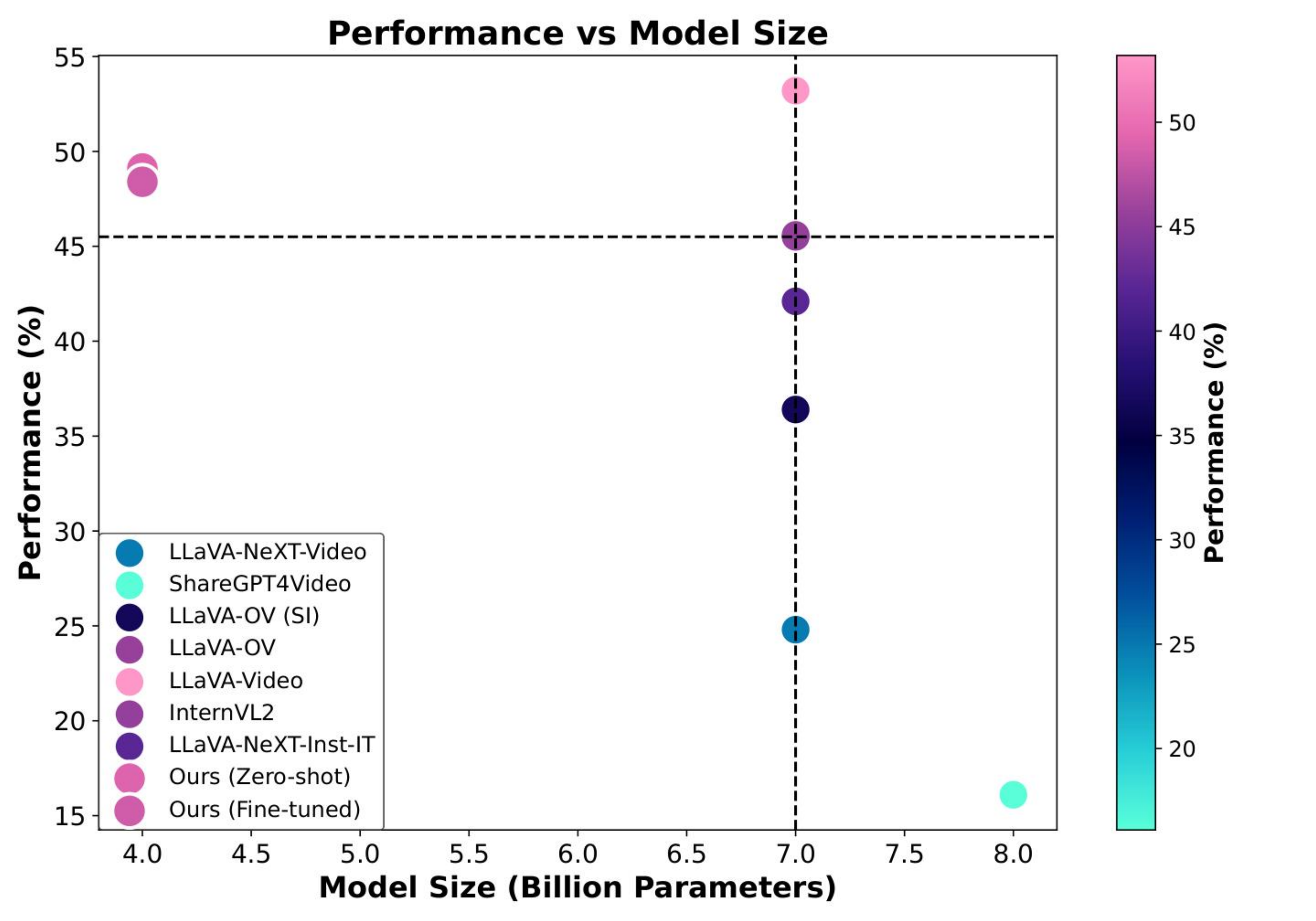}
        \caption{}
    \end{subfigure}
   \caption{Performance on the Inst-IT benchmark. (a) shows that our model remains highly competitive with baseline methods, indicating that training does not degrade its scene understanding on this complex instance-level task. (b) demonstrates the model’s superior size–performance trade-off, with a 4B-parameter model achieving performance comparable to models exceeding 7B parameters.}
    \label{fig:insit}
\end{figure*}

\paragraph{Inst-IT Bench.}
The second dataset, \textit{Inst-IT Bench} \cite{peng2024inst}, is a fine-grained multimodal benchmark designed to evaluate instance-level comprehension in both images and videos. It provides a more detailed assessment of a model’s ability to reason about specific entities and their interactions within a scene.
Formally, each sample in Inst-IT Bench can be represented as:
\begin{equation}
x = (M, Q, \mathcal{A}, a^*),
\end{equation}
where $M$ denotes the input modality (either an image or a video), $Q$ is the query, $\mathcal{A}$ is the set of candidate answers (or a free-form answer space in the open-ended setting), and $a^*$ is the ground-truth answer. Inst-IT Bench consists of two splits:  \textit{Image Split} containing 1,036 question-answer pairs over 338 images, and \textit{Video Split} containing 1,001 question-answer pairs over 206 videos. Each question-answer pair is provided in both multiple-choice and open-ended formats, enabling comprehensive evaluation of both constrained reasoning and generative capabilities. An example of this dataset is provided in \Cref{fig:insit}.

As the main goal of this paper is intention recognition, we train our models on the train set IntentQA dataset and evaluate the trained model on the test and Validation sets of IntentQA plus on the test set of Inst-IT bench.
\begin{table*}[t]
\centering
\begin{tabular}{lc|cc|cc|cc|cc}
\toprule
 \multirow{2}{*}{Model} 
& \multirow{2}{*}{Setting}
& \multicolumn{2}{c|}{ROUGE-1}
& \multicolumn{2}{c|}{ROUGE-L}
& \multicolumn{2}{c|}{CosSim}
& \multicolumn{2}{c}{BERTScore-F1} \\
\cmidrule(lr){3-4} \cmidrule(lr){5-6} \cmidrule(lr){7-8} \cmidrule(lr){9-10}
 & & Val & Test & Val & Test & Val & Test & Val & Test \\
\midrule

 Qwen3-VL 2B & Zero-shot
& 17.18 & 15.60
& 16.88 & 15.30
& 33.35 & 31.13
& 87.78 & 87.33 \\

 Qwen3-VL 2B & Finetuned
& 13.05 & 14.10
& 12.35 & 13.30
& 35.15 & 35.53
& 86.00 & 86.18 \\



\midrule
 Qwen3-VL 4B  & Zero-shot
& 19.71 & 19.75
& 18.95 & 19.00
& 41.07 & 41.04
& 88.68 & 88.55 \\

Qwen3-VL 4B &  Finetuned
& 12.89 & 13.20
& 12.63 & 12.98
& 35.85 & 35.94
& 85.82 & 85.98 \\


\midrule
 \textbf{IntentVLM (Ours)}& \textbf{2-step (ft) }
& \textbf{21.05 }& \textbf{19.18}
& \textbf{20.48} & \textbf{18.91}
& \textbf{35.75 }& \textbf{34.67}
& \textbf{87.75} & \textbf{87.20} \\

\bottomrule
\end{tabular}
\caption{Comparison of intention understanding performance on the IntentQA dataset, evaluated with and without answer options. All values are reported as percentages (\%) and averaged across CW, CH, TP, and TN metrics.}
\label{tab:intentqa_universal}
\end{table*}

\begin{figure}
    \centering
    \includegraphics[width=1\linewidth]{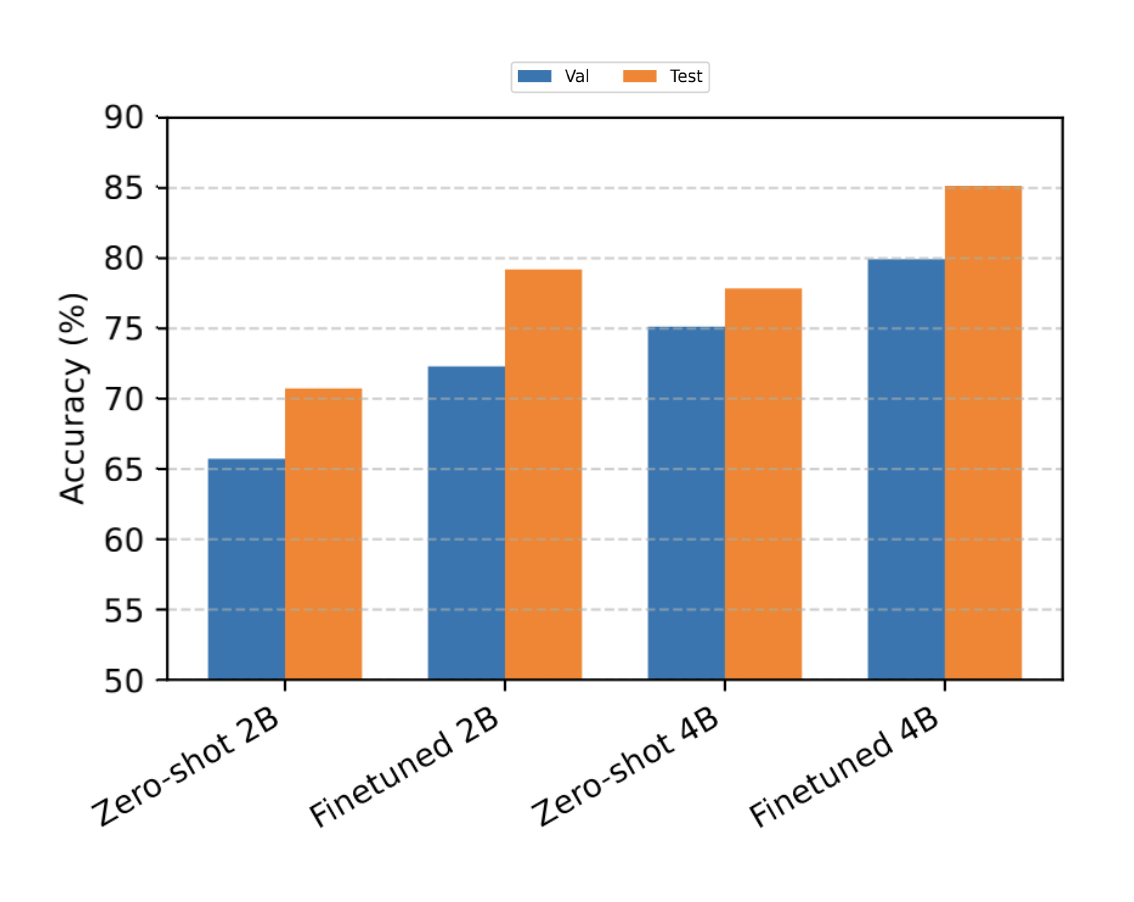}
        \caption{Ablation study analyzing the impact of VLM scale (2B vs.\ 4B) and training regime (zero-shot vs.\ fine-tuned) on performance.}
    \label{fig:abl}
\end{figure}

\subsection{Training Procedure}

For training, we adopt \textit{Qwen3-VL} \cite{bai2025qwen3} as the backbone video-language model and fine-tune it separately for each stage of the proposed framework. To ensure computational efficiency, we employ parameter-efficient fine-tuning (PEFT) \cite{zhang2025parameter} using Low-Rank Adaptation (LoRA) \cite{hu2022lora}, allowing the model to adapt to the task with a minimal number of trainable parameters.

\paragraph{Goal Candidate Proposal.}
In the first stage, the model is trained to generate a set of plausible goal (intention) candidates given the multimodal input. The input consists of the video $V$, the query $Q$, and an instruction $U$ that prompts the model to generate multiple candidate options. The supervision signal is provided by the candidate options available in the IntentQA dataset. Formally, this stage models the conditional distribution over goal candidates: $G \sim F_{\phi}(V, Q, U),$
where $G = \{g_1, g_2, \dots, g_K\}$ denotes the set of generated candidate goals. Here, $F_{\phi}$ is a video--language model parameterized by $\phi$ (implemented via LoRA adapters), $V$ is the input video, $Q$ is the textual query, and $U$ is the instruction prompt.

\paragraph{Intention Selection.}
In the second stage, we fine-tune another instance of the Qwen3-VL backbone to perform intention inference by selecting the most plausible option from the candidate set. The model takes as input the video $V$, the query $Q$, the instruction $U$, and the set of candidate options $G$, and outputs the selected intention. Formally, this stage is defined as: $P_\theta(i \mid U, V, Q, G), \quad \text{where } i \in O$, where $i$ denotes the selected intention (i.e., the most likely goal) from the candidate set $O$.

\subsection{Experimental Setup}

For the intention selection task, we post-train two variants of Qwen3-VL \cite{bai2025qwen3}, namely the 2B and 4B models, using a consistent supervised fine-tuning (SFT) configuration. We adopt a per-device batch size of 1 with gradient accumulation over 8 steps to maintain stable optimization under memory constraints. The models are trained for 2 epochs with a learning rate of $2 \times 10^{-5}$ and a warmup ratio of 0.1. We use the AdamW optimizer implemented in PyTorch, with bfloat16 (bf16) precision enabled and fp16 disabled. For the intention goal proposal task, we employ a larger Qwen3-VL 8B model, as this task requires generating diverse and distinguishable candidate intentions rather than selecting from predefined options. Despite the increased model capacity, we retain the same training configuration as in the selection setting to ensure comparability and stability. The larger model capacity enables improved generative performance, which is critical for producing high-quality and semantically distinct candidate goals.

\subsection{Metrics and Baselines}

For intention detection and instance detection, which are formulated as classification tasks (i.e., selecting the correct answer from a set of candidates), we adopt \textit{accuracy} as the primary evaluation metric. Accuracy measures the proportion of correctly predicted labels over the total number of samples and is standard for multiple-choice video question answering benchmarks.

As baselines for intention recognition, we compare our approach against a diverse set of models reported in the IntentQA \cite{li2023intentqa} benchmark, including \textit{EVQA}, \textit{CoMem}, \textit{HGA}, \textit{HME}, \textit{HQGA}, and \textit{VGT}, as well as \textit{Blind GPT} and \textit{VideoQA}, which are very variant in terms of size and methodology(e.g. LLM-based and Graph-based). The human performance on the benchmark and the details of the experiment are reported in \cite{li2023intentqa}. However, for the Inst-IT benchmark, we use the state of the art video-language models that are in the same category as our models in terms of size. These models include \textit{LLaVA-NeXT-Video, ShareGPT4Video, LLaVA-OV (SI), LLaVA-OV, LLaVA-Video, InternVL2, and LLaVA-NeXT-Inst-IT \cite{peng2024inst}.} 

For the open-vocabulary comparison setting, where no goal candidate are provided and the model must generate responses conditioned on video content, we formulate the task as a free-form video question answering problem. We consider three evaluation settings as baselines: (1) direct answer generation, where the model produces an answer without intermediate reasoning; (2) answer selection, where the model selects from a set of candidate answers; and (3) our proposed approach, which first generates candidate goals (intentions) and then selects the most appropriate one. Each setting is evaluated in both zero-shot and fine-tuned regimes.

To assess the quality of generated answers, we employ complementary automatic metrics. \textit{ROUGE-1} measures unigram overlap, while \textit{ROUGE-L} captures similarity based on the longest common subsequence between generated and reference texts \cite{barbella2022rouge}. \textit{Cosine Similarity (CosSim)} is computed between sentence embeddings to evaluate semantic similarity at the representation level \cite{chandrasekaran2021evolution}. Finally, \textit{BERTScore F1} leverages contextualized embeddings from pretrained language models to measure token-level semantic alignment, providing robustness beyond surface-form matching \cite{hanna2021fine}. 

\section{Results}

As shown in \Cref{tab:main_results}, our experiments on the IntentVQA benchmark demonstrate that the proposed selection model built on a 4B parameter version of Qwen3-VL achieves substantial improvements over baseline methods. Specifically, the model attains an average accuracy of approximately 80\% across both validation and test sets, representing a relative improvement of around 30\% compared to existing baselines. Notably, the model also matches human performance, highlighting its effectiveness in intention understanding tasks. \Cref{fig:abl} presents an ablation study evaluating the impact of model scale (4B vs.\ 2B parameters) and training regime (fine-tuned vs.\ zero-shot). The results show that increasing model capacity consistently improves performance. Furthermore, fine-tuning on IntentVQA training examples significantly boosts accuracy compared to zero-shot settings, particularly when the model is required to select from predefined options. This suggests that exposure to task-specific data enhances the model's ability to generalize to both validation and test scenarios. These findings demonstrate the strong capability and potential of forward-inverse modeling with video-language models for complex reasoning tasks such as intention detection.

However, as shown in \Cref{tab:intentqa_universal}, performance degrades substantially in the open-vocabulary setting where candidates are not provided. In this scenario, the zero-shot model exhibits suboptimal behavior, achieving less than 20 points in terms of ROUGE score, despite maintaining relatively high contextual similarity as measured by BERTScore F1. Furthermore, fine-tuning the model on direct-answer supervision does not alleviate this issue and, in fact, leads to a decrease across all evaluated metrics. Intention options typically introduce constraints, disambiguation, and additional context that effectively reduce the solution space, enabling the model to exploit structural shortcuts. When these options are removed and the model is fine-tuned solely on question–answer pairs, inference becomes an unconstrained generation task with a substantially larger output space. This mismatch in task formulation degrades performance, as the model is not adequately trained to operate in such open-ended settings. In contrast, applying our proposed two-stage method yields a consistent improvement, increasing performance by approximately 5 points on average.

On the other hand, as shown in \Cref{fig:insit}(a), our experiments on the InsIT benchmark indicate that training on the IntentQA dataset for intention recognition does not lead to catastrophic forgetting. The performance of the fine-tuned model remains very close to its zero-shot counterpart, suggesting that task-specific adaptation preserves generalization ability. Furthermore, as illustrated in \Cref{fig:insit}(b), both models achieve competitive performance compared to models specifically trained on the InsIT benchmark, despite being approximately half their size. This highlights the efficiency of our approach in maintaining strong cross-task performance while using comparatively smaller models. 

These results establish structured intention recognition as a key component for enabling video-language foundation models to support more socially aware and cognitively capable robotic systems.

\section*{Safe and Responsible Innovation Statement}
 While our system is designed to improve human-robot collaboration, inferring human intent from video introduces privacy risks, as continuous monitoring may be perceived as surveillance. We acknowledge that models trained on limited public egocentric datasets may reflect cultural biases, potentially underperforming for underrepresented populations. Deployment in assistive or caregiving contexts demands particular care to avoid misinterpretation of vulnerable individuals' behavior. 

\section*{Conclusion}
This paper introduced \textit{IntentVLM}, a novel two-stage video-language framework designed for open-vocabulary human intention recognition. The approach is inspired by forward-inverse modeling in cognitive science by decomposing intention understanding into goal candidate generation followed by structured inference through selection. Experimental results on IntentQA and Inst-IT Bench confirm that IntentVLM significantly outperforms existing baselines and matches human-level performance while maintaining strong generalization across external tasks. Our work demonstrates that structured reasoning enables high-level cognitive inference in multimodal models. 

\bibliographystyle{ieeetr}
\bibliography{reference}

@article{Baker2009Action,
    author    = {Baker, Chris L. and Saxe, Rebecca and Tenenbaum, Joshua B.},
    title     = {Action understanding as inverse planning},
    journal   = {Cognition},
    volume    = {113},
    number    = {3},
    pages     = {329--349},
    year      = {2009},
    publisher = {Elsevier},
    doi       = {10.1016/j.cognition.2009.07.005}
}

@inproceedings{trick2019multimodal,
  title={Multimodal uncertainty reduction for intention recognition in human-robot interaction},
  author={Trick, Susanne and Koert, Dorothea and Peters, Jan and Rothkopf, Constantin A},
  booktitle={2019 IEEE/RSJ International Conference on Intelligent Robots and Systems (IROS)},
  pages={7009--7016},
  year={2019},
  organization={IEEE}
}

@article{deRuiter2013,
  title={Forward modelling requires intention recognition and non-impoverished predictions},
  author={de Ruiter, Jan P. and Cummins, Chris},
  journal={Behavioral and Brain Sciences},
  volume={36},
  number={4},
  pages={393--394},
  year={2013}
}

@article{Baker2009,
  title={Action understanding as inverse planning},
  author={Baker, Chris L. and Saxe, Rebecca and Tenenbaum, Joshua B.},
  journal={Cognition},
  volume={113},
  number={3},
  pages={329--349},
  year={2009}
}

@article{Qian2021,
  title={Modeling human intention inference from visual motion using inverse planning},
  author={Qian, Zhutian and Kryven, Mykhailo and Gao, Tianshu and Tenenbaum, Joshua B.},
  journal={arXiv preprint arXiv:2112.00903},
  year={2021}
}

@inproceedings{Ng2000,
  title={Algorithms for inverse reinforcement learning},
  author={Ng, Andrew Y. and Russell, Stuart J.},
  booktitle={Proceedings of the 17th International Conference on Machine Learning (ICML)},
  pages={663--670},
  year={2000}
}

@inproceedings{HadfieldMenell2016,
  title={Cooperative inverse reinforcement learning},
  author={Hadfield-Menell, Dylan and Russell, Stuart J. and Abbeel, Pieter and Dragan, Anca D.},
  booktitle={Advances in Neural Information Processing Systems (NeurIPS)},
  pages={3909--3917},
  year={2016}
}

@article{zhang2025parameter,
  title={Parameter-efficient fine-tuning for foundation models},
  author={Zhang, Dan and Feng, Tao and Xue, Lilong and Wang, Yuandong and Dong, Yuxiao and Tang, Jie},
  journal={arXiv preprint arXiv:2501.13787},
  year={2025}
}

@inproceedings{hanna2021fine,
  title={A fine-grained analysis of BERTScore},
  author={Hanna, Michael and Bojar, Ond{\v{r}}ej},
  booktitle={Proceedings of the Sixth Conference on Machine Translation},
  pages={507--517},
  year={2021}
}

@article{chandrasekaran2021evolution,
  title={Evolution of semantic similarity—a survey},
  author={Chandrasekaran, Dhivya and Mago, Vijay},
  journal={Acm Computing Surveys (Csur)},
  volume={54},
  number={2},
  pages={1--37},
  year={2021},
  publisher={ACM New York, NY, USA}
}

@article{barbella2022rouge,
  title={Rouge metric evaluation for text summarization techniques},
  author={Barbella, Marcello and Tortora, Genoveffa},
  journal={Available at SSRN 4120317},
  year={2022}
}

@article{hu2022lora,
  title={Lora: Low-rank adaptation of large language models.},
  author={Hu, Edward J and Shen, Yelong and Wallis, Phillip and Allen-Zhu, Zeyuan and Li, Yuanzhi and Wang, Shean and Wang, Liang and Chen, Weizhu and others},
  journal={Iclr},
  volume={1},
  number={2},
  pages={3},
  year={2022}
}

@article{bai2025qwen3,
  title={Qwen3-vl technical report},
  author={Bai, Shuai and Cai, Yuxuan and Chen, Ruizhe and Chen, Keqin and Chen, Xionghui and Cheng, Zesen and Deng, Lianghao and Ding, Wei and Gao, Chang and Ge, Chunjiang and others},
  journal={arXiv preprint arXiv:2511.21631},
  year={2025}
}

@article{HoGriffiths2022,
  title={Cognitive science as a source of forward and inverse models of human decision making},
  author={Ho, Mark K. and Griffiths, Thomas L.},
  journal={Psychological Review},
  year={2022}
}

@article{peng2024inst,
  title={INST-IT: Boosting Instance Understanding via Explicit Visual Prompt Instruction Tuning},
  author={Peng, Wujian and Meng, Lingchen and Chen, Yitong and Xie, Yiweng and Liu, Yang and Gui, Tao and Xu, Hang and Qiu, Xipeng and Wu, Zuxuan and Jiang, Yu-Gang},
  journal={arXiv preprint arXiv:2412.03565},
  year={2024}
}

@inproceedings{li2023intentqa,
  title={Intentqa: Context-aware video intent reasoning},
  author={Li, Jiapeng and Wei, Ping and Han, Wenjuan and Fan, Lifeng},
  booktitle={Proceedings of the IEEE/CVF international conference on computer vision},
  pages={11963--11974},
  year={2023}
}

@article{ho2022cognitive,
  title={Cognitive science as a source of forward and inverse models of human decisions for robotics and control},
  author={Ho, Mark K and Griffiths, Thomas L},
  journal={Annual Review of Control, Robotics, and Autonomous Systems},
  volume={5},
  number={1},
  pages={33--53},
  year={2022},
  publisher={Annual Reviews}
}

@inproceedings{nguyen2011capir,
  title={Capir: Collaborative action planning with intention recognition},
  author={Nguyen, Truong-Huy and Hsu, David and Lee, Wee-Sun and Leong, Tze-Yun and Kaelbling, Leslie and Lozano-Perez, Tomas and Grant, Andrew},
  booktitle={Proceedings of the AAAI Conference on Artificial Intelligence and Interactive Digital Entertainment},
  volume={7},
  number={1},
  pages={61--66},
  year={2011}
}

@article{jain2019probabilistic,
  title={Probabilistic human intent recognition for shared autonomy in assistive robotics},
  author={Jain, Siddarth and Argall, Brenna},
  journal={ACM Transactions on Human-Robot Interaction (THRI)},
  volume={9},
  number={1},
  pages={1--23},
  year={2019},
  publisher={ACM New York, NY, USA}
}

@article{brune2007mental,
  title={Mental state attribution, neurocognitive functioning, and psychopathology: what predicts poor social competence in schizophrenia best?},
  author={Br{\"u}ne, Martin and Abdel-Hamid, Mona and Lehmk{\"a}mper, Caroline and Sonntag, Claudia},
  journal={Schizophrenia research},
  volume={92},
  number={1-3},
  pages={151--159},
  year={2007},
  publisher={Elsevier}
}

@article{ziemke2020understanding,
  title={Understanding robots},
  author={Ziemke, Tom},
  journal={Science Robotics},
  volume={5},
  number={46},
  pages={eabe2987},
  year={2020},
  publisher={American Association for the Advancement of Science}
}

@article{thellman2022mental,
  title={Mental state attribution to robots: A systematic review of conceptions, methods, and findings},
  author={Thellman, Sam and De Graaf, Maartje and Ziemke, Tom},
  journal={ACM Transactions on Human-Robot Interaction (THRI)},
  volume={11},
  number={4},
  pages={1--51},
  year={2022},
  publisher={ACM New York, NY}
}

@Inbook{Chetouani2026,
author="Chetouani, Mohamed",
editor="Chetouani, Mohamed
and Nowak, Andrzej
and Lukowicz, Paul",
title="Introduction to Computational Human-AI Collaboration",
bookTitle="Handbook of Human-AI Collaboration",
year="2026",
publisher="Springer Nature Switzerland",
address="Cham",
isbn="978-3-031-61050-9"
}

@article{hoffman2024inferring,
  title={Inferring human intent and predicting human action in human--robot collaboration},
  author={Hoffman, Guy and Bhattacharjee, Tapomayukh and Nikolaidis, Stefanos},
  journal={Annual Review of Control, Robotics, and Autonomous Systems},
  volume={7},
  number={1},
  pages={73--95},
  year={2024},
  publisher={Annual Reviews}
}

@article{dreyfus2007,
    author    = {Hubert L. Dreyfus},
    title     = {Detachment, involvement, and rationality: Are we essentially rational animals?},
    journal   = {Human Affairs},
    volume    = {17},
    number    = {2},
    pages     = {101--109},
    year      = {2007},
    doi = {10.2478/v10023-007-0010-0}
}

@article{buyukgoz2021proactive,
    author = {Buyukgoz, S. and Grosinger J. and Chetouani, M. and Saffiotti, A.},
    title = {Two ways to make your robot proactive: Reasoning about human intentions or reasoning about possible futures},
    journal = {Frontiers in Robotics and AI},
    year = {2022},
    doi = {10.3389/frobt.2022.929267}
}

@article{van2024proactive,
  title={What is proactive human-robot interaction?-a review of a progressive field and its definitions},
  author={van Den Broek and Marike Koch and Moeslund, Thomas B},
  journal={ACM Transactions on Human-Robot Interaction},
  volume={13},
  number={4},
  pages={1--30},
  year={2024},
  publisher={ACM New York, NY}
}

@inproceedings{Lian2024OVMER,
author = {Lian, Zheng and Sun, Haiyang and Sun, Licai and Chen, Haoyu and Chen, Lan and Gu, Hao and Wen, Zhuofan and Chen, Shun and Zhang, Siyuan and Yao, Hailiang and Liu, Bin and Liu, Rui and Liang, Shan and Li, Ya and Yi, Jiangyan and Tao, Jianhua},
title = {OV-MER: towards open-vocabulary multimodal emotion recognition},
year = {2025},
publisher = {JMLR.org},
booktitle = {Proceedings of the 42nd International Conference on Machine Learning},
articleno = {1466},
numpages = {36},
series = {ICML'25}
}

@inproceedings{Schrempf2007Tractable,
  author    = {Schrempf, Oliver C. and Albrecht, D. and Hanebeck, Uwe D.},
  title     = {Tractable probabilistic models for intention recognition based on expert knowledge},
  booktitle = {2007 {IEEE/RSJ} International Conference on Intelligent Robots and Systems},
  pages     = {3122--3127},
  year      = {2007},
  doi       = {10.1109/IROS.2007.4399226}
}

@article{VanHorenbeke2021Activity,
  author    = {Van-Horenbeke, F. A. and Peer, A.},
  title     = {Activity, plan, and goal recognition: A review},
  journal   = {Frontiers in Robotics and {AI}},
  volume    = {8},
  year      = {2021},
  doi       = {10.3389/frobt.2021.643010}
}

@inproceedings{huang2022inner,
    title={Inner Monologue: Embodied Reasoning through Planning with Language Models},
    author={Wenlong Huang and Fei Xia and Ted Xiao and Harris Chan and Jacky Liang and Pete Florence and Andy Zeng and Jonathan Tompson and Igor Mordatch and Yevgen Chebotar and Pierre Sermanet and Tomas Jackson and Noah Brown and Linda Luu and Sergey Levine and Karol Hausman and brian ichter},
    booktitle={6th Annual Conference on Robot Learning},
    year={2022}
}

@article{huang2022icml,
    author = {Wenlong Huang and Pieter Abbeel and Deepak Pathak and Igor Mordatch},
    title = {Language models as zero-shot planners: Extracting actionable knowledge for embodied agents},
    journal = {International Conference on Machine Learning},
    year = {2022}
}

@article{khan2025fmreview,
    author = {Muhammad Tayyab Khan and Ammar Waheed},
    title = {Foundation Model Driven Robotics: A Comprehensive Review},
    journal = {arXiv preprint arXiv:2507.10087v1},
    year = {2025}
}

@article{clip,
    author = {Alec Radford and Jong Wook Kim and Chris Hallacy and Aditya Ramesh and Gabriel Goh and Sandhini Agarwal and Girish Sastry and Amanda Askell and Pamela Mishkin and Jack Clark and Gretchen Krueger and Ilya Sutskever},
    title = {Learning transferable visual models from natural language supervision},
    booktitle = {International Conference on Machine Learning},
    year = {2021}
}

@inproceedings{Kautz1986plan,
author = {Kautz, Henry A. and Allen, James F.},
title = {Generalized plan recognition},
year = {1986},
publisher = {AAAI Press},
booktitle = {Proceedings of the Fifth AAAI National Conference on Artificial Intelligence},
pages = {32–37},
numpages = {6},
series = {AAAI'86}
}

@article{li2021visualnlp,
    author = {Li Z and Mu Y and Sun Z and Song S and Su J and Zhang J},
    title = {Intention Understanding in Human–Robot Interaction Based on Visual-NLP Semantics},
    journal = {Frontiers in Neurobotics},
    doi = {10.3389/fnbot.2020.610139},
    year = {2021}
}

@inproceedings{chiu2021,
    author = {Yu-Ching Chiu and Bo-Hao Chang and Tzu-Yu Chen and Cheng-Fu Yang},
    title = {Multi-modal User Intent Classification Under the Scenario of Smart Factory},
    booktitle = {Proceedings of the AAAI Conference on Artificial Intelligence},
    doi = {10.1609/aaai.v35i18.17882},
    year = {2021}
}

@inproceedings{sap2022tom,
    title = {Neural Theory-of-Mind? On the Limits of Social Intelligence in Large {LM}s},
    author = {Sap, Maarten  and Le Bras, Ronan  and Fried, Daniel  and Choi, Yejin},
    editor = {Goldberg, Yoav  and Kozareva, Zornitsa and Zhang, Yue},
    booktitle = {Proceedings of the 2022 Conference on Empirical Methods in Natural Language Processing},
    year = {2022},
    publisher = {Association for Computational Linguistics},
    doi = {10.18653/v1/2022.emnlp-main.248},
    pages = {3762--3780},
}

@article{nageris2024draco,
    author = {Ben Nageris and Felipe Meneguzzi and Reuth Mirsky},
    title = {Goal Recognition using Actor-Critic Optimization},
    journal = {arXiv preprint arXiv:2501.01463},
    year = {2024}
}

@article{masters2019,
author = {Masters, Peta and Sardina, Sebastian},
title = {Cost-based goal recognition in navigational domains},
year = {2019},
volume = {64},
number = {1},
issn = {1076-9757},
doi = {10.1613/jair.1.11343},
journal = {J. Artif. Int. Res.},
pages = {197–242}
}

@book{sukthankar2014pair,
    author = {Gita Sukthankar and Christopher Geib and Hung Hai Bui and David Pynadath and Robert P Goldman},
    title = {Plan, Activity, and Intent Recognition: Theory and Practice},
    publisher = {Morgan Kaufmann},
    year = {2014}
}

@article{bratman1989,
    author = {Michael E. Bratman},
    title = {Intention and personal policies},
    journal = {Philosophical Perspectives},
    doi = {10.2307/2214277},
    year = {1989}
}

@inproceedings{charniak1991pr,
author = {Charniak, Eugene and Goldman, Robert},
title = {A probabilistic model of plan recognition},
year = {1991},
isbn = {0262510596},
publisher = {AAAI Press},
booktitle = {Proceedings of the Ninth National Conference on Artificial Intelligence - Volume 1},
pages = {160–165},
numpages = {6},
series = {AAAI'91}
}

@inproceedings{Ramrez2009PlanRA,
author = {Ram\'{\i}rez, Miquel and Geffner, Hector},
title = {Plan recognition as planning},
year = {2009},
publisher = {Morgan Kaufmann Publishers Inc.},
booktitle = {Proceedings of the 21st International Joint Conference on Artificial Intelligence},
pages = {1778–1783},
numpages = {6},
location = {Pasadena, California, USA},
series = {IJCAI'09}
}

@inproceedings{ramrez2010pr,
author = {Ram\'{\i}rez, Miquel and Geffner, Hector},
title = {Probabilistic plan recognition using off-the-shelf classical planners},
year = {2010},
publisher = {AAAI Press},
booktitle = {Proceedings of the Twenty-Fourth AAAI Conference on Artificial Intelligence},
pages = {1121–1126},
numpages = {6},
series = {AAAI'10}
}

@article{wei2022llm,
title={Emergent Abilities of Large Language Models},
author={Jason Wei and Yi Tay and Rishi Bommasani and Colin Raffel and Barret Zoph and Sebastian Borgeaud and Dani Yogatama and Maarten Bosma and Denny Zhou and Donald Metzler and Ed H. Chi and Tatsunori Hashimoto and Oriol Vinyals and Percy Liang and Jeff Dean and William Fedus},
journal={Transactions on Machine Learning Research},
issn={2835-8856},
year={2022},
url={https://openreview.net/forum?id=yzkSU5zdwD},
note={Survey Certification}
}

@inproceedings{brown2020fewshotlearners,
author = {Brown, Tom B. and Mann, Benjamin and Ryder, Nick and Subbiah, Melanie and Kaplan, Jared and Dhariwal, Prafulla and Neelakantan, Arvind and Shyam, Pranav and Sastry, Girish and Askell, Amanda and Agarwal, Sandhini and Herbert-Voss, Ariel and Krueger, Gretchen and Henighan, Tom and Child, Rewon and Ramesh, Aditya and Ziegler, Daniel M. and Wu, Jeffrey and Winter, Clemens and Hesse, Christopher and Chen, Mark and Sigler, Eric and Litwin, Mateusz and Gray, Scott and Chess, Benjamin and Clark, Jack and Berner, Christopher and McCandlish, Sam and Radford, Alec and Sutskever, Ilya and Amodei, Dario},
title = {Language models are few-shot learners},
year = {2020},
isbn = {9781713829546},
publisher = {Curran Associates Inc.},
booktitle = {Proceedings of the 34th International Conference on Neural Information Processing Systems},
articleno = {159},
numpages = {25},
series = {NIPS '20}
}

@article{madan2024foundation,
  title={Foundation models for video understanding: A survey},
  author={Madan, Neelu and M{\o}gelmose, Andreas and Modi, Rajat and Rawat, Yogesh S and Moeslund, Thomas B},
  journal={arXiv preprint arXiv:2405.03770},
  year={2024}
}

@article{zhang2024vision,
  title={Vision-language models for vision tasks: A survey},
  author={Zhang, Jingyi and Huang, Jiaxing and Jin, Sheng and Lu, Shijian},
  journal={IEEE transactions on pattern analysis and machine intelligence},
  volume={46},
  number={8},
  pages={5625--5644},
  year={2024},
  publisher={IEEE}
}

@inproceedings{gao2024vlm,
  author={Gao, Jensen and Sarkar, Bidipta and Xia, Fei and Xiao, Ted and Wu, Jiajun and Ichter, Brian and Majumdar, Anirudha and Sadigh, Dorsa},
  booktitle={2024 IEEE International Conference on Robotics and Automation (ICRA)}, 
  title={Physically Grounded Vision-Language Models for Robotic Manipulation}, 
  year={2024},
  pages={12462-12469},
  doi={10.1109/ICRA57147.2024.10610090}
}

@inproceedings{
lin2025reinforcement,
title={Reinforcement Learning for Human-{AI} Collaboration via Probabilistic Intent Inference},
author={Yuxin Lin and Seyede Fatemeh Ghoreishi and Tian Lan and Mahdi Imani},
booktitle={Reinforcement Learning Conference},
year={2025},
url={https://openreview.net/forum?id=u5bi4lzEYx}
}

@inproceedings{wan2025fiser,
author = {Wan, Yanming and Wu, Yue and Wang, Yiping and Mao, Jiayuan and Jaques, Natasha},
title = {Infer human's intentions before following natural language instructions},
year = {2025},
isbn = {978-1-57735-897-8},
publisher = {AAAI Press},
url = {https://doi.org/10.1609/aaai.v39i24.34718},
doi = {10.1609/aaai.v39i24.34718},
articleno = {2820},
numpages = {9},
series = {AAAI'25/IAAI'25/EAAI'25}
}

@inproceedings{ying2024goma,
  author={Ying, Lance and Jha, Kunal and Aarya, Shivam and Tenenbaum, Joshua B. and Torralba, Antonio and Shu, Tianmin},
  booktitle={2024 IEEE/RSJ International Conference on Intelligent Robots and Systems (IROS)}, 
  title={GOMA: Proactive Embodied Cooperative Communication via Goal-Oriented Mental Alignment}, 
  year={2024},
  pages={7099-7106},
  doi={10.1109/IROS58592.2024.10802144}
}

@inproceedings{zhang2025combo,
  author={Hongxin Zhang and Zeyuan Wang and Qiushi Lyu and Zheyuan Zhang and Sunli Chen and Tianmin Shu and Behzad Dariush and Kwonjoon Lee and Yilun Du and Chuang Gan},
  title={COMBO: Compositional World Models for Embodied Multi-Agent Cooperation},
  year={2025},
  cdate={1735689600000},
  url={https://openreview.net/forum?id=YXRyYkb1im},
  booktitle={ICLR},
}

@inproceedings{lesh1995gr,
author = {Lesh, Neal and Etzioni, Oren},
title = {A sound and fast goal recognizer},
year = {1995},
isbn = {1558603638},
publisher = {Morgan Kaufmann Publishers Inc.},
booktitle = {Proceedings of the 14th International Joint Conference on Artificial Intelligence - Volume 2},
pages = {1704–1710},
numpages = {7},
series = {IJCAI'95}
}

@INPROCEEDINGS{li2025vlmreview,
  author={Li, Zongxia and Wu, Xiyang and Du, Hongyang and Liu, Fuxiao and Nghiem, Huy and Shi, Guangyao},
  booktitle={2025 IEEE/CVF Conference on Computer Vision and Pattern Recognition Workshops (CVPRW)}, 
  title={A Survey of State of the Art Large Vision Language Models: Alignment, Benchmark, Evaluations and Challenges}, 
  year={2025},
  pages={1578-1597},
  doi={10.1109/CVPRW67362.2025.00147}}

@article{huang2024lit,
    author = {Zhe Huang and John Pohovey and Ananya Yammanuru and Katherine Driggs-Campbell},
    title = {LIT: Large Language Model Driven Intention Tracking for Proactive Human-Robot Collaboration -- A Robot Sous-Chef Application},
    journal = {arXiv preprint arXiv:2406.13787},
    year = {2024}
}

@article{zhao2025deep,
  title={Deep learning approaches for multimodal intent recognition: A survey},
  author={Zhao, Jingwei and Wen, Yuhua and Li, Qifei and Hu, Minchi and Zhou, Yingying and Xue, Jingyao and Wu, Junyang and Gao, Yingming and Wen, Zhengqi and Tao, Jianhua and others},
  journal={arXiv preprint arXiv:2507.22934},
  year={2025}
}

@inproceedings{rahimi2025user,
  title={User-vlm: Llm contextualization with multimodal pre-trained user models},
  author={Rahimi, Hamed and Abrini, Mouad and Khoramshahi, Mahdi and Chetouani, Mohamed},
  booktitle={ToM4AI@ 39th Annual AAAI Conference on Artificial Intelligence},
  year={2025}
}

@inproceedings{rahimi2025user2,
  title={User-vlm 360: Personalized vision language models with user-aware tuning for social human-robot interactions},
  author={Rahimi, Hamed and Bahaj, Adil and Abrini, Mouad and Khoramshahi, Mahdi and Ghogho, Mounir and Chetouani, Mohamed},
  booktitle={Proceedings of the 27th International Conference on Multimodal Interaction},
  pages={326--336},
  year={2025}
}

@article{malecot2026harmoni,
  title={HARMONI: Multimodal Personalization of Multi-User Human-Robot Interactions with LLMs},
  author={Mal{\'e}cot, Jeanne and Rahimi, Hamed and Cattoni, Jeanne and Samson, Marie and Abrini, Mouad and Khoramshahi, Mahdi and Pino, Maribel and Chetouani, Mohamed},
  journal={arXiv preprint arXiv:2601.19839},
  year={2026}
}

@article{grislain2025failsense,
  title={I-FailSense: Towards General Robotic Failure Detection with Vision-Language Models},
  author={Grislain, Clemence and Rahimi, Hamed and Sigaud, Olivier and Chetouani, Mohamed},
  journal={arXiv preprint arXiv:2509.16072},
  year={2025}
}

@inproceedings{rahimi2025reasoning,
  title={Reasoning llms for user-aware multimodal conversational agents},
  author={Rahimi, Hamed and Cattoni, Jeanne and Beghili, Meriem and Abrini, Mouad and Khoramshahi, Mahdi and Pino, Maribel and Chetouani, Mohamed},
  booktitle={2025 34th IEEE International Conference on Robot and Human Interactive Communication (RO-MAN)},
  pages={443--448},
  year={2025},
  organization={IEEE}
}

@article{cretides2026encoding,
  title={Encoding Predictability and Legibility for Style-Conditioned Diffusion Policy},
  author={Cr{\'e}tides, Adrien Jacquet and Abrini, Mouad and Rahimi, Hamed and Chetouani, Mohamed},
  journal={arXiv preprint arXiv:2603.16368},
  year={2026}
}

@inproceedings{rahimi2025demographic,
  title={Demographic User Modeling for Social Robotics with Multimodal Pre-trained Models},
  author={Rahimi, Hamed and Abrini, Mouad and Malecot, Jeanne and Lai, Ying and Jacquet Cr{\'e}tides, Adrien and Khoramshahi, Mahdi and Chetouani, Mohamed},
  booktitle={Proceedings of the 27th International Conference on Multimodal Interaction},
  pages={337--343},
  year={2025}
}

@article{abrini2025proceedings,
  title={Proceedings of 1st Workshop on Advancing Artificial Intelligence through Theory of Mind},
  author={Abrini, Mouad and Abend, Omri and Acklin, Dina and Admoni, Henny and Aichinger, Gregor and Alon, Nitay and Ashktorab, Zahra and Atreja, Ashish and Auron, Moises and Aufreiter, Alexander and others},
  journal={arXiv preprint arXiv:2505.03770},
  year={2025}
}

\end{document}